# Population changes in residential clusters in Japan

Takuya Sekiguchi[1,2], Kohei Tamura[3], Naoki Masuda[4*]

1. National Institute of Informatics, 2-1-2 Hitotsubashi, Chiyoda-ku, Tokyo 101-8430, Japan
2. JST, ERATO, Kawarabayashi Large Graph Project, c/o Global Research Center for Big Data Mathematics, NII, Chiyoda-ku, Tokyo, Japan
3. Frontier Research Institute for Interdisciplinary Sciences, Tohoku University, 6-3 Aramaki aza Aoba, Aoba-ku, Sendai 980-8578, Japan
4. Department of Engineering Mathematics, University of Bristol, Merchant Venturers Building, Woodland Road, Clifton, Bristol BS8 1UB, United Kingdom

* Author for correspondence: naoki.masuda@bristol.ac.uk

## Abstract

Population dynamics in urban and rural areas are different. Understanding factors that contribute to local population changes has various socioeconomic and political implications. In the present study, we use population census data in Japan to examine contributors to the population growth of residential clusters between years 2005 and 2010. The data set covers the entirety of Japan and has a high spatial resolution of 500 × 500 m$^2$, enabling us to examine population dynamics in various parts of the country (urban and rural) using statistical analysis. We found that, in addition to the area, population density, and age, the shape of the cluster and the spatial distribution of inhabitants within the cluster are significantly related to the population growth rate of a residential cluster. Specifically, the population tends to grow if the cluster is "round" shaped (given the area) and the population is concentrate near the center rather than periphery of the cluster.

## Keywords

urban; population census; spatial patterns; city clustering algorithm; residential cluster

## 1. Introduction

Population change is a central precondition to be considered in policy making and urban planning. In urban areas with high population concentrations, decentralization policies may be designed to mitigate congestion and environmental problems [1]. In developing countries, rapid growth of the number of urban dwellers is forecasted to exacerbate water shortage [2]. In rural areas facing population aging and shrinkage, how to ensure convenience of public transportation [3] and health care services [4] is a crucial issue.

The choice of the residential location is a main determinant of spatial patterns of population changes over time. People have been suggested to choose the residential location by considering residential environment attributes such as the accessibility to workplace measured by commute distance [5–7], school quality [8, 9], and the crime rate [8, 10]. Residential mobility is also affected by the individual's life course and household attributes such as age and income [7, 10], job change [5], marital status [11], the numbers of children and drivers [10], and home ownership [7, 11].

In addition to these factors, spatial characteristics of the city and inhabited areas, which shape socioeconomic and geographical environments, may also impact spatio-temporal patterns of population changes. For example, urban sprawl is considered to be a consequence of uncoordinated and unplanned urban development [12] and results in scattered spatial patterns of employment and residences in suburban areas [13–16]. These spatial patterns would cause a long commute time due to poor accessibility to workplaces [17]. In contrast, compact urban growth and the diversity of land uses within the region enhance the accessibility to both work and non-work activities [18, 19]. If the accessibility to workplaces and other activities influences residential decision-making, spatial patterns of inhabited regions are expected to affect dynamics of population changes.

There have been studies relating the population size or its change to spatial patterns of urban areas. For example, the population size of a region was shown to obey a power-law relationship with the area of the region in Norfolk in England [20] (also see [21] for an analysis of approximately 70,000 cities in the world). In 78 regions in Israel, the population growth rate in sprawl regions was higher than in compact regions, where the sprawl and compact regions were defined in part by the shape of their boundaries [22]. Fractal dimensions are also useful tools for relating the population size/growth and spatial patterns of residential areas. For example, the fractal dimension of the central part of Tel

Aviv metropolis and its population size concomitantly increased over time, and the observed fractal dimension was larger than that of the wider Tel Aviv [23]. In 20 urban areas in the US, the fractal dimension and the population size were positively correlated [24].

To the best of our knowledge, past studies on the relationship between spatial characteristics of regions and population changes examined a single or a small number of metropolitan areas of interest. Therefore, it seems to be unknown whether the relationship between spatial characteristics of regions and population changes can be generalized to a large number of metropolitan and non-metropolitan areas, even within a country. To address this question, one needs longitudinal data of population density with a high spatial resolution. Remote sensing technologies and the recent prevalence of mobile phones offer promising data on population dynamics at relatively low cost [25–27]. For example, the spatial distribution of the number of workers estimated from mobile phone data closely matched the counterpart calculated from the US census data [28]. The population density can also be estimated from the amount of night-time lights in satellite imagery [29, 30]. Such data enable estimation of short-term human mobility within a day or week [31, 32].

However, the accuracy of data obtained with these technologies is unclear. Furthermore, the population dynamics estimated by these methods may be susceptible to changes in the accuracy and coverage of the technology over time. In the present study, we use population census data of Japan with a high spatial resolution measured five years apart. To date, census data are probably advantageous to mobile phone or satellite data in tracking long-term population changes with a high accuracy. In fact, census data have been used for evaluating the accuracy of other techniques [28, 29].

We explore spatial factors that contribute to the population growth in local clusters of inhabited areas. We hypothesize that the shape of the cluster of inhabited patches significantly affects the population change in the cluster. To test the hypothesis, we carry out statistical analysis to relate population changes in a cluster over five years, from 2005 to 2010, to the cluster’s shape and other demographic and socioeconomic variables. We resolve the aforementioned limitations of the previous studies by exhaustively analyzing clusters of inhabited areas across Japan and by using the census data with which the local populations are accurately estimated.

## 2. Methods

## 2.1 Data set

We used data obtained from the population census of Japan in 2005 and 2010; the census is conducted every five years. The data consist of demographic information on a grid of cells of 500 m × 500 m covering the entire Japan [33–35]. There are 1,944,711 cells in total including completely water-surface cells (e.g., sea and lake), of which 482,181 cells were populated in 2005 and 477,172 cells in 2010. The population was 127,767,994 in 2005 (65,419,017 females and 62,348,977 males) and 128,057,352 in 2010 (65,729,615 females and 62,327,737 males). The numbers of female inhabitants, that of male inhabitants, and the latitude and longitude of the center of the cell are available for each cell. We denote the year (i.e., 2005 or 2010) by $t$.

## 2.2 City clustering algorithm

To determine the boundary of an inhabited area, we applied the city clustering algorithm [21, 36–38]. The algorithm calculates the connected components of populated cells, i.e., cells that contain at least one inhabitant, where we have defined the adjacency of cells by the von Neumann neighborhood (i.e., each cell has four neighbors in the north, south, east, and west). We refer to each connected component as cluster and discuss how its properties affect the population changes of the cluster over the five years, i.e., between 2005 and 2010. We obtained 24,165 and 24,707 clusters in 2005 and 2010, respectively.

## 2.3 Dependent variable

We denote by $n_i(t)$ the number of inhabitants in cell $i$ at time $t$. To investigate dynamics of the population within a cluster, we adopted regression models whose dependent variable is

$$\tilde{n}_c^{\text{cluster}}(2010) = \sum_{i\in\text{cluster } c} n_i(2010), \qquad (1)$$

i.e., the number of inhabitants in 2010 in cluster $c$ observed in 2005 (Fig. 1). With this definition, we aimed to examine population dynamics in the clusters that existed in 2005. We used $\log(n_c^{\text{cluster}}(2005))$, where

$$n_c^{\text{cluster}}(2005) = \sum_{i\in\text{cluster } c} n_i(2005), \qquad (2)$$

i.e., the number of inhabitants in cluster $c$ in 2005, as the offset variable (see Eq. (7)).

Cells in a cluster observed in 2005 may belong to different clusters recalculated in 2010. Furthermore, some inhabited cells in 2010 do not belong to any cluster observed in 2005 (Fig. 1). Reflecting the latter fact, although the total population of Japan in 2005 is equal to $\sum_c n_c^{\text{cluster}}(2005) = 127{,}767{,}994$, the sum $\sum_c \tilde{n}_c^{\text{cluster}}(2010) = 127{,}901{,}037$ is smaller than the total population of Japan in 2010.

### 2.4 Independent variables

We used the following independent variables for each cluster observed in 2005 to explain the population change between 2005 and 2010.

First, the area of the cluster (denoted by $S$ and referred to as *Area*) is defined by the number of cells constituting the cluster. Second, the population density (referred to as *Density*) is equal to the number of inhabitants in the cluster divided by $S$.

We quantified the shape of the cluster by the following two indices. We defined what we refer to as *Roundness*, originally proposed in Ref. [39], as $S$ divided by the area of the circle whose diameter is equal to the longest Euclidean distance between two cells belonging to the cluster. We measured the position of a cell by the two-dimensional coordinate of the center of the cell. For example, the clusters shown in Fig. 2 have three cells and have the longest Euclidean distance equal to two (in the unit of the linear length of a cell), yielding a *Roundness* value of 0.955. A cluster whose shape is close to a circle yields a large *Roundness* value. For a given $S$, the line-shaped cluster yields the smallest *Roundness* value. *Roundness* can be regarded as a simplified variant of the box-counting fractal dimension [40]. The second shape-related index, *Irregularity*, is defined by

$$\frac{2\log L}{\log S}, \qquad (3)$$

where $L$ is the perimeter of the cluster. For a fixed $S$, *Irregularity* is small when the cluster is close to square-shaped. The perimeter was used for characterizing spatial patterns of urban regions [20]. Frenkel and Ashkenazi [22] applied Eq. (3) to quantify the level of urban sprawl. We note that measures similar to *Irregularity* were proposed decades ago [41, 42].

We quantified the hypothesized efficiency of communication or transportation within a cluster by the following two indices. We defined the expected distance between uniformly randomly selected two inhabitants in the cluster by

$$\frac{\sum_{i \le j: i,j \in \text{cluster } c} n_i(2005) n_j(2005) d_{ij}}{\binom{n_c^{\text{cluster}}(2005)}{2}}, \qquad (4)$$

where $d_{ij}$ is the distance between cells $i$ and $j$, and the denominator of Eq. (4) is a binomial coefficient. It should be noted that $n_i(2005)n_j(2005)/\binom{n_c^{\text{cluster}}(2005)}{2}$ is the probability that randomly selected two inhabitants in cluster $c$ belong to cells $i$ and $j$. Because Eq. (4) has the dimension of the length, it may give rise to multicollinearity with $S$ in multivariate regression. To mitigate this potential problem, we divided Eq. (4) by $\sqrt{S}$, which has a dimension of the length, to define

$$\frac{\sum_{i\le j:i,j\in\text{cluster } c} n_i(2005)n_j(2005)d_{ij}}{\binom{n_c^{\text{cluster}}(2005)}{2}}/\sqrt{S}\,. \qquad (5)$$

Equation (5) would be dimensionless if clusters are two-dimensional (with a large *Roundness* and/or small *Irregularity* value). In fact, clusters may be line-shaped or fractal-like, in which case Eq. (5) would have a dimension of the length to some power. However, we expect that Eq. (5) is less correlated with $S$ than Eq. (4) is. Therefore, we adopted Eq. (5) as a dependent variable and referred to it as *Distance*. We also adopted the coefficient of variation, which is defined by the standard deviation divided by the mean, of the number of inhabitants in a cell belonging to the focal cluster. This index quantifies spatial heterogeneity in the distribution of inhabitants within the cell and is referred to as *Heterogeneity*.

Figure 2 illustrates the difference among *Density*, *Distance*, and *Heterogeneity*. The three clusters shown in the figure have the same *Area* (= 3) and *Density* (= 2.333). However, *Heterogeneity* for the clusters shown in Figs. 2(a) and 2(b) (= 0.990) is larger than that for the cluster shown in Fig. 2(c) (= 0.495). *Distance* is smaller for the cluster shown in Fig. 2(b) (= 0.330) than that shown in Fig. 2(a) (= 0.440), because in Fig. 2(b) the most populated cell is located in the center of the cluster. Note that the distribution of the number of inhabitants in a cluster is the same between Figs. 2(a) and 2(b). *Distance* is the largest for the cluster shown in Fig. 2(c) (= 0.495).

We used the following two demographic dependent variables. First, *Gender* refers to the fraction of female inhabitants in the cluster. Second, we estimated the average age of the inhabitants in a cluster, referred to as *Age*, as follows. Because the data set did not contain the average age for each cell, we approximated it by the average age of inhabitants in the prefecture to which a cluster belongs. The average age of inhabitants in each prefecture is available from the prefacer-level population census data carried out in 2005 [43]. The prefecture of a cluster was defined as the prefecture to which the cell with the largest closeness centrality [44, 45] in the cluster belongs. In the calculation of the closeness centrality, we regarded the cluster as a network in which a cell was a node and two nodes were adjacent if they shared a side. Using the R package ‘ggmap’[46], we sent the latitude and longitude of the cell with the largest closeness centrality to Google Map API and detected the prefecture. If multiple cells had the same largest closeness centrality value, we used the average latitude and longitude of these cells to determine the cluster’s prefecture.

As a socioeconomic factor, we used the fraction of workers in the tertiary industry in the prefecture to which the cluster belongs [43] and referred to it as *Tertiary*. We determined the prefecture of a cluster in the same manner as in the case of *Age*.

### 2.5 Regression models

For analysis of count data, a Poisson regression model is often used (e.g., [47]). This model assumes that the dependent variable ($\tilde{n}_c^{\text{cluster}}(2010)$ in the present case) obeys a Poisson distribution given by

$$\Pr\left(\tilde{n}_c^{\text{cluster}}(2010) = k\right) = \frac{\exp(-\mu_c)\mu_c^k}{k!}, \qquad (6)$$

where the conditional mean $\mu_c$ is determined by

$$\log(\mu_c) = \log\left(n_c^{\text{cluster}}(2005)\right) + \beta_0 + \beta_1 \log(Area_c) + \beta_2 \log(Density_c) + \sum_{i=3}^{9} \beta_i X_{i,c}. \quad (7)$$

In Eq. (7), Eq. (2) is used as the offset variable, the logarithmic link function is used, $\beta_0$ is the intercept, $\beta_i$ ($i$ = 1,…, 9) is a regression coefficient, $X_i$ ($i$ = 3,…, 9) is the $i$th independent variable (i.e., *Roundness*, *Irregularity*, *Distance*, *Heterogeneity*, *Gender*, *Age*, and *Tertiary*), and subscript $c$ on the right-hand side indicates that the values of the independent variables are for cluster $c$.

In the Poisson regression model, the conditional mean of the dependent variable is assumed to be equal to its conditional variance. However, as we will show in Section 3.1, the conditional variance of the dependent variable is considerably larger than its conditional mean for the present data. This situation is called the overdispersion, which we tested by running an overdispersion test [48, 49] (see also [50] for the usage of R package ‘AER’). The overdispersion test is carried out based on the statistic

$$\frac{\sum_{c=1}^{N}\left(\left(\tilde{n}_c^{\text{cluster}}(2010)-\hat{\mu}_c\right)^2-\tilde{n}_c^{\text{cluster}}(2010)\right)}{\sqrt{2\sum_{c=1}^{N}\hat{\mu}_c{}^2}}, \qquad (8)$$

which asymptotically obeys the normal distribution with mean 0 and standard deviation 1 under the assumption of the Poisson model. In Eq. (8), $\hat{\mu}_c$ is the maximum likelihood estimate of the dependent variable under the Poisson model (i.e., the null hypothesis).

Because the null hypothesis was rejected (Section 3.1), we used the negative binomial regression model. A negative binomial regression model [47] assumes that the dependent variable obeys a negative binomial distribution given by

$$\Pr\left(\tilde{n}_c^{\text{cluster}}(2010)=k\right)=\frac{\Gamma(k+\theta)}{\Gamma(\theta)\Gamma(k+1)}\left(\frac{\theta}{\mu_c+\theta}\right)^{\theta}\left(\frac{\mu_c}{\mu_c+\theta}\right)^{k}, \qquad (9)$$

where $\Gamma(\cdot)$ is the gamma function, and $\theta$ is a parameter that is assumed to be the same for all clusters. In Eq. (9), the conditional mean, $\mu_c$, is given by Eq. (7). The variance of the distribution given by Eq. (9) is $\mu_c+\mu_c^2/\theta$. To fit the model, we maximized the likelihood with respect to $\beta_i$ ($i = 0, \ldots, 9$) (Eq. (7)) and $\theta$ using glm.nb() function in R package ‘MASS’ [51].

In Eq. (7), we logarithmically transformed *Area* and *Density* to improve linearity between the dependent and independent variables. In fact, they obeyed long-tailed distributions (see Section 3.1). For these two independent variables, a 1% increase in *Area* (or *Density*) corresponds to a $\beta_1$ (or $\beta_2$) % increase in the number of inhabitants in 2010 in a cluster observed in 2005. For $X_i$ ($i = 3, \ldots, 9$), an increase in $X_i$ by one unit increases the number of inhabitants $\exp(\beta_i)$ times. We used the same offset term Eq. (2) in the multivariate and univariate regressions.

We also searched the multivariate regression model that minimized the Akaike information criterion (AIC) among the models that had any of the independent variables as main effects and any of pairwise interaction terms between the independent variables. To avoid large variance inflation factor (VIF) values due to the pairwise interaction terms, we normalized all independent variables to have a zero mean [52]. We used the stepwise backward elimination method to find the best model, i.e., by sequentially excluding the least significant term in terms of the AIC [53].

# 3. Results

## 3.1 Descriptive statistics

Statistics of the dependent, offset, and independent variables are shown in Table 1. We find that the area of a cluster, $S$, the number of inhabitants in a cluster, and the population density in a cluster are heterogeneously distributed, as suggested by large coefficient of variation (CV) values for these variables. This observation is confirmed by long-tailed distributions of these quantities shown in Fig. 3.

In the following statistical analysis, we restricted ourselves to the clusters whose areas are at least ten cells because the geometry of smaller clusters would be strongly affected by the spatial discreteness. We ran the overdispersion test to confirm that the assumption of the Poisson distribution of the dependent variable was violated ($p < 0.001$). Therefore, in the following we report the result of the negative binomial regression model.

### 3.2 Correlation coefficients

The Pearson, Spearman, and Kendall correlation coefficients between pairs of independent variables are shown in Tables 2(a), 2(b), and 2(c), respectively. The signs of almost all of the correlation coefficients are consistent across the three types of correlation coefficient.

Table 2(a) indicates that log(*Area*) and *Irregularity* are strongly correlated (Pearson correlation coefficient = −0.794). However, we concluded that the multicollinearity problem was not present because the VIF values were sufficiently small (4.206 and 3.247 for log(*Area*) and *Irregularity*, respectively). In general, VIF values for independent variables should be less than 10, preferably less than 5, for multivariate regression analysis to be justified [54, 55].

### 3.3 Regression analysis

The results of the negative binomial regression are shown in Table 3. The contributions of log(*Area*) and log(*Density*) were significant at the 0.1% level, *Irregularity* and *Age* at the 1% level, and *Distance* at the 10% level. The other variables (i.e., *Roundness*, *Heterogeneity*, *Gender*, and *Tertiary*) were not significant. Table 3 also indicates that a 1% increase in *Area* and *Density* is associated with an increase in the number of inhabitants in a cluster in 2010 (as compared to 2005) by 0.0113% and 0.0227%, respectively. An increase in *Irregularity*, *Age*, and *Distance* by 1% is associated with a decrease in the number of inhabitants in a cluster by $3.27\times10^{-4}$ ($=1-\exp(-0.0327\times0.01)$) times, $2.50 \times 10^{-5}$ ($= 1-\exp(-0.0025 \times 0.01)$) times, and $3.62 \times 10^{-4}$ ($= 1-\exp(-0.0362\times0.01)$) times, respectively. Because the total population in Japan only changed by 0.23% between 2005 and 2010 (Section 2.1), the contribution of these factors to the population change is non-negligible.

The results for univariate regressions are also shown in Table 3. The signs of all the significant regression coefficients in the multivariate regression (i.e., negative binomial regression) were consistent with the results for the univariate regression, lending support to the results obtained from the multivariate analysis.

We carried out the model selection in terms of the Akaike Information Criterion (AIC) among the negative binomial regression models that were allowed to include any main effects and pairwise interaction terms. The regression coefficients of the selected model are shown in Table 4. The selected model contained all independent variables. The result that the main effects of log(*Area*), log(*Density*), and *Distance* are significant is consistent with that for the multivariate regression. However, the main effects of *Irregularity* and *Age*, which were significant in the multivariate regression, were not significant in the selected model, while some interaction effects between other variables and *Irregularity* or *Age* were significant. This result implies that the effects of *Irregularity* and *Age* qualitatively depend on other variables. Lastly, the main effect of *Heterogeneity*, which was not significant in the multivariate regression, was significant in the selected model.

On the basis of the results for the multivariate regression, univariate regression, and model selection, we conclude that the main effects of *Area*, *Density*, and *Distance* are significant according to the different criteria. In other words, the population growth of a cluster is associated with an increase in *Area*, an increase in *Density*, and a decrease in *Distance*. In addition, the main effects of *Irregularity* and *Age* were also significant in the multivariate and univariate regression (but not in the model selected by the AIC).

## 4. Discussion

We searched for potential drivers of population changes in terms of demographic, geometrical, and other properties of a cluster of inhabited cells. Unsurprisingly, we found that the area and the population density of the cluster were positively correlated to the population growth rate. In addition, we found that a shape parameter for the cluster, *Irregularity*, and the mean distance between inhabitants within the cluster, *Distance*, had negative impacts on the population growth. *Age* also had a negative impact on the population growth. In contrast, the fraction of female inhabitants, *Gender*, and that of tertiary-industry workers, *Tertiary*, had no significant contribution. The present results suggest that the population change is predictable to a certain degree from spatial characteristics intrinsic to the cluster, irrespectively of demographic factors.

The result that the *Irregularity* variable is negatively correlated with the population growth rate (in the multivariate and univariate regressions) implies that spatial compactness of a cluster would lead to an increase in a cluster's population. Because urban sprawl is often negatively associated with the compact city [17, 22], it is intriguing to associate urban sprawl with *Irregularity*. However, urban sprawl is not solely characterized by the shape of urban areas but also by a discontinuous development of suburban areas, which may reduce the intra- and inter-region accessibility [13]. To relate our approach to urban sprawl, we probably need to consider relationships between different clusters and the role of each cluster in wider geographical regions.

*Distance* had a negative impact on the population growth rate. By definition, *Distance* is small when highly populated cells are located near the geographical center of a cluster (Fig. 2(b)) rather than when they are located in the periphery of the cluster (Fig. 2(a)). Therefore, our results suggest that a cluster's population tends to grow if many inhabitants are located near the center of the cluster. A previous study showed that the values of indices characterizing urban regions (e.g., Moran, Geary, and Gini coefficients) were sensitive to the distribution of inhabitants in a confined region [56]. The present study suggests that the spatial distribution of inhabitants may affect the population growth rate as well as such urban indices. Investigating this issue warrants future work.

We did not pay attention to the change in the shape of the cluster over years. In fact, processes of urban growth, which are characterized by, for example, the population size, economic performance, and development of transportation systems, occur in tandem with changes in the shape of urban areas (e.g., [23, 57, 58]). Socioeconomic factors reflected in the shape of urban areas may influence inhabitants' residential decision making, which may in turn change the shape of urban areas.

An important limitation of the present study is that we did not have an access to migration data. In general, the population change is decomposed into the natural increase (i.e., births minus deaths) and the migratory increase (i.e., immigration minus emigration). Because the census data used in the present study did not include the information about the population flow, we could not distinguish between the natural and migratory increases. Another limitation is that some dependent variables (i.e., *Age* and *Tertiary*) were estimated at the prefecture level due to the lack of data at the level of single cells. We did not consider other information such as land use as independent variables, either. Inclusion of any of these variables with an appropriate spatial resolution will be an interesting extension of the present work.

# Declarations

**Ethics approval and consent to participate**

Not applicable

**Consent for publication**

Not applicable

**Availability of data and material**

The population census data for each cell in 2005 and 2010 are available at Ref. [33] and Ref. [34], respectively. The average age of inhabitants and the fraction of workers in the tertiary industry in each prefecture of Japan are available in Tables No. 19 and No. 31 in Ref. [43], respectively.

**Competing interests**

The authors declare that they have no competing interests.

**Funding**

This work was supported by JST ERATO Grant Number JPMJER1201, Japan, and JST CREST Grant Number JPMJCR1304.

**Authors' contributions**

NM designed the study. KT collected the data. TS analyzed the data. All authors wrote the manuscript.

## Figures

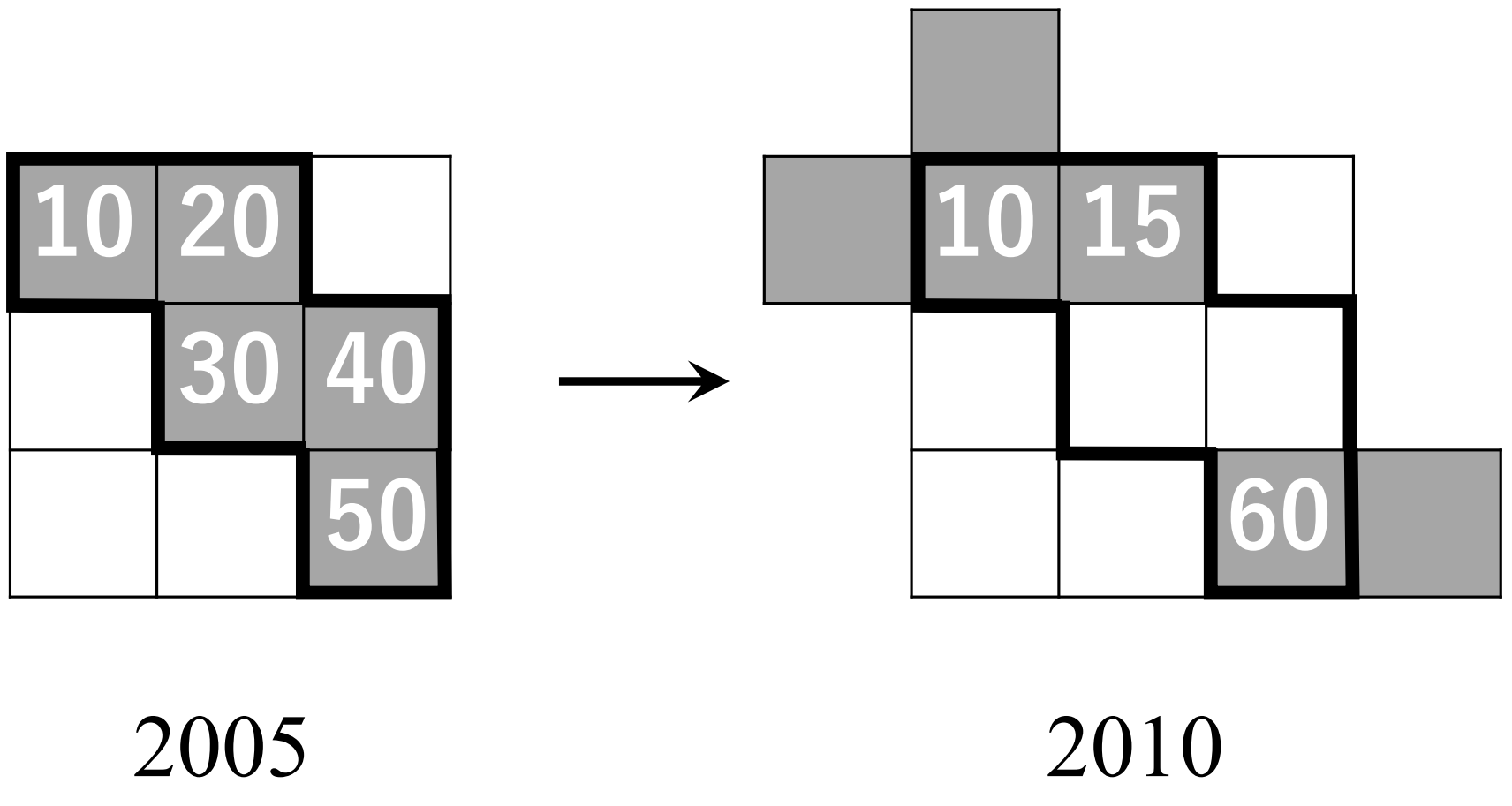

**Figure 1: A hypothetical example of the population change in a cluster over five years.** The number of inhabitants in a cell is indicated for inhabited cells shown in gray. The bold lines indicate the boundary of cluster *c* observed in year 2005. This cluster has $n_c^{\text{cluster}}(2005) = \sum_{i \in \text{cluster } c} n_i(2005) = 10 + 20 + 30 + 40 + 50 = 150$ and $\tilde{n}_c^{\text{cluster}}(2010) = \sum_{i \in \text{cluster } c} n_i(2010) = 10 + 15 + 60 = 85$ inhabitants in 2005 and 2010, respectively. Although cluster *c* is split into different clusters in 2010, each of which extends beyond the border of cluster *c* determined in 2005, we neglect the split to calculate the population change in cluster *c*. Therefore, cluster *c* has lost 150 – 85 = 65 inhabitants in the five years.

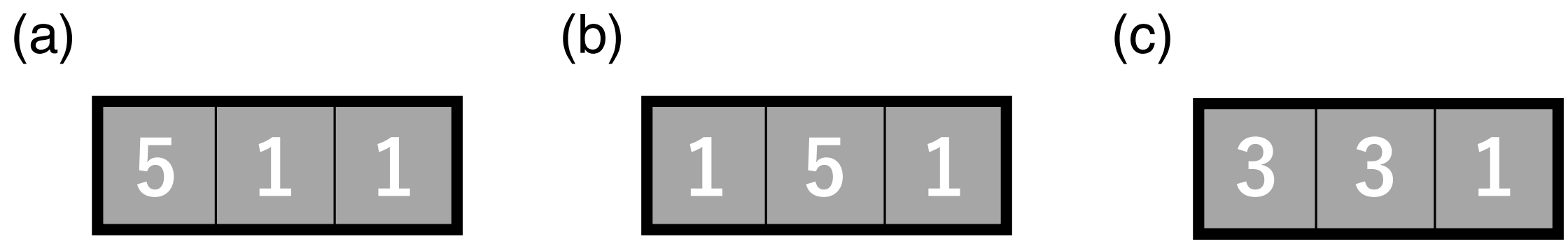

**Figure 2: Three clusters, each composed of three cells (i.e., the area of the cluster, *S* = 3) and seven inhabitants.** The number of inhabitants in each cell is indicated.

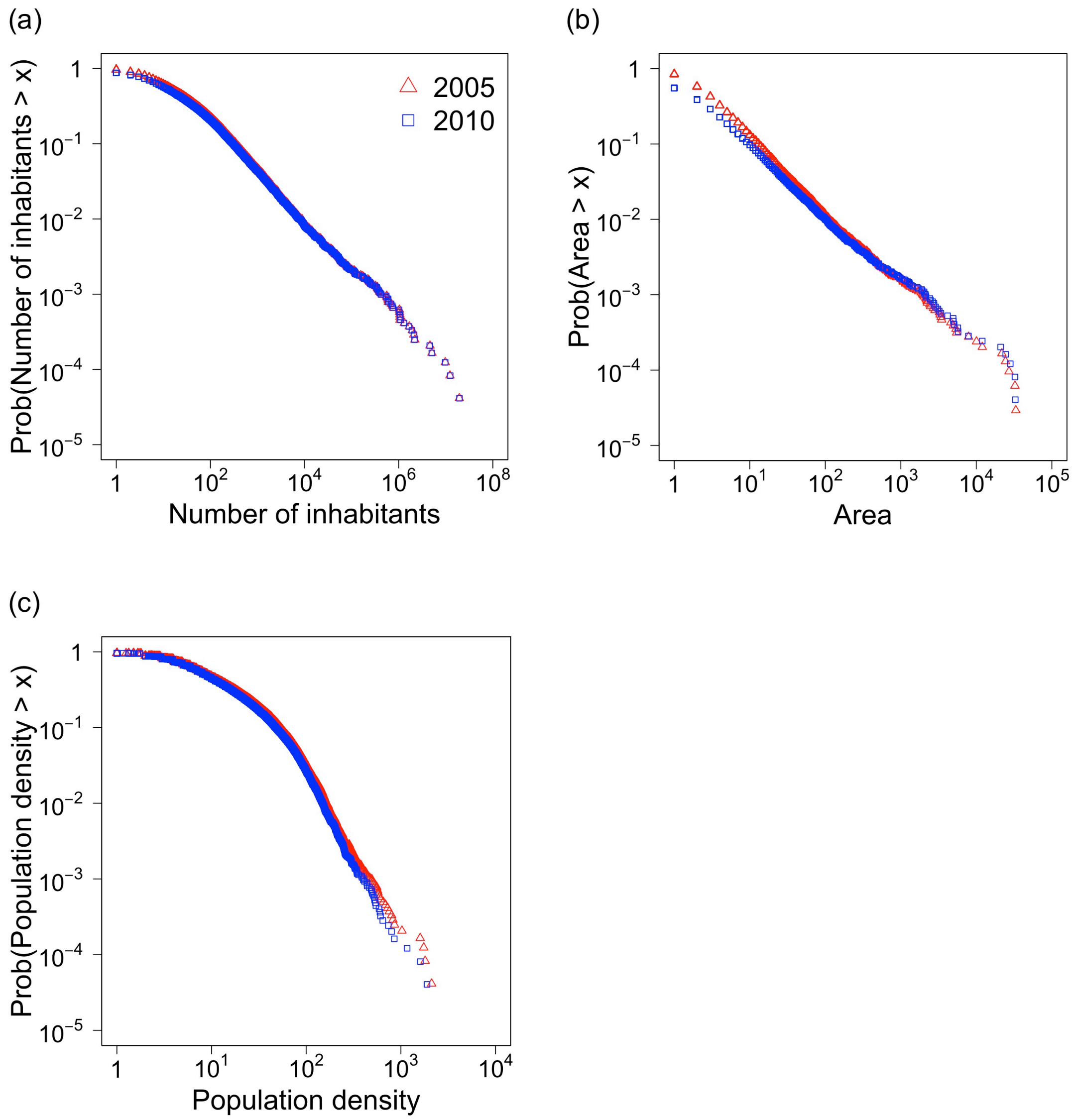

**Figure 3: Complementary cumulative distributions for three properties of a cluster.** **(a)** Number of inhabitants. **(b)** Area. **(c)** Population density.

**Table 1: Descriptive statistics for the clusters composed of at least ten cells.** There are $N = 2689$ clusters. By definition, all independent variables are calculated for the data in 2005. SD represents the standard deviation. CV represents the coefficient of variation. The non-standardized *Area* and *Density* variables are not used in the regression models but shown for completeness.

| | Min | Max | Median | Mean | SD | CV |
|---|---|---|---|---|---|---|
| **Dependent variable** | | | | | | |
| $\tilde{n}_c^{\text{cluster}}(2010)$ | 22 | 42,492,718 | 623 | 47,193.57 | 964,925.07 | 20.446 |
| **Offset** | | | | | | |
| $n_c^{\text{cluster}}(2005)$ | 39 | 41,403,322 | 696 | 47,086.07 | 946,332.04 | 20.098 |
| **Independent variables** | | | | | | |
| log(*Area*) | 2.303 | 11.187 | 2.944 | 3.256 | 1.070 | 0.329 |
| log(*Density*) | 1.291 | 7.684 | 3.509 | 3.526 | 0.977 | 0.277 |
| *Roundness* | 0.066 | 1.273 | 0.397 | 0.434 | 0.200 | 0.460 |
| *Irregularity* | 1.746 | 2.685 | 2.336 | 2.327 | 0.181 | 0.078 |
| *Distance* | 0.110 | 0.720 | 0.296 | 0.301 | 0.078 | 0.260 |
| *Heterogeneity* | 0.342 | 3.190 | 1.052 | 1.119 | 0.387 | 0.346 |
| *Gender* | 0.236 | 0.694 | 0.523 | 0.520 | 0.027 | 0.051 |
| *Age* | 39.1 | 47.1 | 44.4 | 44.6 | 1.240 | 0.028 |
| *Tertiary* | 0.570 | 0.774 | 0.653 | 0.653 | 0.045 | 0.070 |
| *Area* | 10 | 72,194 | 19 | 160.50 | 1,896.4 | 11.816 |
| *Density* | 3.636 | 2,172.5 | 33.40 | 56.19 | 86.59 | 1.541 |

**Table 2: Correlation coefficient between the independent variables for the clusters with at least ten cells observed in 2005 ($N$ = 2689). (a)** Pearson. **(b)** Spearman. **(c)** Kendall. $^{\dagger}p < 0.1$; $^{**}p < 0.01$; $^{***}p < 0.001$. The VIF of each independent variable is also shown in (a).

(a) Pearson

| | log(*Area*) | log(*Density*) | *Roundness* | *Irregularity* | *Distance* | *Heterogeneity* | *Gender* | *Age* | VIF |
|---|---|---|---|---|---|---|---|---|---|
| log(*Area*) | – | | | | | | | | 4.206 |
| log(*Density*) | 0.431$^{***}$ | – | | | | | | | 1.537 |
| *Roundness* | –0.479$^{***}$ | –0.167$^{***}$ | – | | | | | | 2.163 |
| *Irregularity* | –0.794$^{***}$ | –0.494$^{***}$ | 0.230$^{***}$ | – | | | | | 3.247 |
| *Distance* | 0.232$^{**}$ | –0.132$^{***}$ | –0.594$^{***}$ | 0.009 | – | | | | 2.353 |
| *Heterogeneity* | 0.516$^{***}$ | 0.361$^{***}$ | –0.215$^{***}$ | –0.460$^{***}$ | –0.265$^{***}$ | – | | | 1.944 |
| *Gender* | 0.062 | 0.172$^{**}$ | –0.051$^{**}$ | –0.067$^{***}$ | 0.003 | 0.056$^{**}$ | – | | 1.066 |
| *Age* | –0.001$^{**}$ | –0.067$^{***}$ | –0.013 | 0.010 | 0.008 | 0.008 | 0.162$^{***}$ | – | 1.153 |
| *Tertiary* | –0.003 | –0.112 | 0.041$^{**}$ | –0.007 | –0.134$^{***}$ | 0.150$^{***}$ | –0.068$^{***}$ | –0.291$^{***}$ | 1.191 |

(b) Spearman

| | log(*Area*) | log(*Density*) | *Roundness* | *Irregularity* | *Distance* | *Heterogeneity* | *Gender* | *Age* |
|---|---|---|---|---|---|---|---|---|
| log(*Area*) | – | | | | | | | |
| log(*Density*) | 0.373*** | – | | | | | | |
| *Roundness* | –0.560*** | –0.204*** | – | | | | | |
| *Irregularity* | –0.870*** | –0.456*** | 0.273*** | – | | | | |
| *Distance* | 0.219*** | –0.173*** | –0.616*** | 0.049** | – | | | |
| *Heterogeneity* | 0.473*** | 0.437*** | –0.218*** | –0.466*** | –0.307*** | – | | |
| *Gender* | 0.084*** | 0.192*** | –0.047** | –0.089*** | –0.022 | 0.101*** | – | |
| *Age* | 0.025 | –0.020 | –0.003 | –0.024 | –0.012 | 0.046** | 0.164*** | – |
| *Tertiary* | 0.003 | –0.114*** | 0.035† | –0.003 | –0.114*** | 0.093*** | –0.020 | –0.224*** |

(c) Kendall

| | log(*Area*) | log(*Density*) | *Roundness* | *Irregularity* | *Distance* | *Heterogeneity* | *Gender* | *Age* |
|---|---|---|---|---|---|---|---|---|
| log(*Area*) | – | | | | | | | |
| log(*Density*) | 0.260*** | – | | | | | | |
| *Roundness* | –0.395*** | –0.138*** | – | | | | | |
| *Irregularity* | –0.700*** | –0.318*** | 0.181*** | – | | | | |
| *Distance* | 0.151*** | –0.115*** | –0.446*** | 0.033** | – | | | |
| *Heterogeneity* | 0.333*** | 0.297*** | –0.146*** | –0.323*** | –0.210*** | – | | |
| *Gender* | 0.057*** | 0.129*** | –0.032** | –0.058*** | –0.014 | 0.067*** | – | |
| *Age* | 0.018 | –0.015 | –0.002 | –0.017 | –0.007 | 0.029** | 0.111*** | – |
| *Tertiary* | 0.002 | –0.079*** | 0.025† | –0.003 | –0.079*** | 0.064*** | –0.010 | –0.171*** |

**Table 3: Coefficients of multivariate and univariate negative binomial regressions ($N = 2689$).** CI: 95% confidence interval. $^{\dagger}p < 0.1$; $^{**}p < 0.01$; $^{***}p < 0.001$.

| | Multivariable | | Univariate | |
|---|---|---|---|---|
| | Estimate | CI | Estimate | CI |
| (Intercept) | –0.0098*** | [–0.1370, 0.1174] | – | – |
| log(*Area*) | 0.0113*** | [0.0076, 0.0150] | 0.0202*** | [0.0182, 0.0222] |
| log(*Density*) | 0.0227*** | [0.0197, 0.0256] | 0.0320*** | [0.0294, 0.0347] |
| *Roundness* | 0.0040 | [–0.0128, 0.0208] | –0.0414*** | [–0.0549, –0.0279] |
| *Irregularity* | –0.0327** | [–0.0553, –0.0102] | –0.1428*** | [–0.1564, –0.1292] |
| *Distance* | –0.0362† | [–0.0763, 0.0037] | –0.0302† | [–0.1078, –0.0879] |
| *Heterogeneity* | –0.0006 | [–0.0083, 0.0071] | 0.0375*** | [0.0310, 0.0441] |
| *Gender* | –0.0420 | [–0.1429, 0.0592] | 0.0821 | [–0.0309, 0.1951] |
| *Age* | –0.0025** | [–0.0044, –0.0006] | –0.0039*** | [–0.0059, –0.0019] |
| *Tertiary* | –0.0116 | [–0.0669, 0.0436] | 0.0353*** | [–0.0235, 0.0942] |

**Table 4: The selected model.** AIC = 29,061. $^{*}p < 0.05$; $^{**}p < 0.01$; $^{***}p < 0.001$.

| | Estimate | | Estimate |
|---|---|---|---|
| (Intercept) | $-0.1202^{***}$ | log(*Area*) × log(*Density*) | $-0.0035^{**}$ |
| log(*Area*) | $0.0205^{***}$ | log(*Area*) × *Gender* | $-0.4738^{***}$ |
| log(*Density*) | $0.0230^{***}$ | log(*Area*) × *Age* | $0.0022^{*}$ |
| *Roundness* | 0.0068 | log(*Area*) × *Tertiary* | $0.0828^{**}$ |
| *Irregularity* | −0.0127 | *Density* × *Gender* | $-0.1234^{*}$ |
| *Distance* | $-0.0661^{**}$ | *Density* × *Age* | −0.0019 |
| *Heterogeneity* | $-0.0099^{*}$ | *Density* × *Tertiary* | −0.0522 |
| *Gender* | −0.0580 | *Roundness* × *Gender* | $-1.0901^{***}$ |
| *Age* | −0.0009 | *Irregularity* × *Gender* | $-1.2133^{*}$ |
| *Tertiary* | −0.0486 | *Distance* × *Age* | $-0.0319^{*}$ |
| | | *Distance* × *Tertiary* | $-1.2001^{***}$ |
| | | *Age* × *Tertiary* | $-0.0637^{***}$ |